
\font\titlefont = cmr10 scaled \magstep2
\magnification=\magstep1
\vsize=20truecm
\voffset=1.75truecm
\hsize=14truecm
\hoffset=1.75truecm
\baselineskip=20pt

\settabs 18 \columns

\def\b{\bigskip}
\def\bb{\bigskip\bigskip}

\def\ce{\centerline}

\def\no{\noindent}

 \rightline{AMES-HET- 94-02}
\rightline{March 7, 1994}
\rightline{(revised version)}
\rightline{hep-ph/9404208}
\bb

\b
\ce{\titlefont{Top Quark As A Topological Color Soliton}}
\b
\bb
\ce{\bf{X. Zhang
   }}

\ce{\it{ Department of Physics and Astronomy}}
\ce{\it{Iowa State University}}
\ce{\it{Ames, Iowa 50011 }}
\bb
\bb
\ce{\bf Abstract}
\b

\no In this paper I propose an scenario for the top quark as a
topological color soliton. I illustrate some features of the top
quark as a soliton in a toy model based on
nonlinear realization of the $SU_L(3) \times SU_R(3) / SU_{L+R}(3)$
 with $SU_{L+R}(3)$ being the color gauge group.

 \filbreak

Recently the CDF collaboration[1] presented evidence for
the top quark with mass $m_t \sim 174$ GeV. Since $m_t$ is of
order the Fermi scale, the top
quark is expected to hold the clue to the physics
of the electroweak symmetry breaking[2].
 The scale associated with top quark mass
generation can
be estimated by imposing unitarity on the processes
$t {\overline{t}} \rightarrow V_L V_L$[3], where
$V_L$ is a longitudinal $W^\pm$ or Z. In the absence
 of the Higgs boson, such an amplitude is proportional to
$G_F m_t \sqrt{s}$, and will violate unitarity at high
energy $\Lambda$.
The largest zeroth-partial-wave amplitude is
the color-singlet, spin-zero combination of $t {\overline{t}}$
and the zero-isospin combination of $W^+_L~ W^-_L$
and
$Z_L ~Z_L$, from which
 one has[4]
$${
\Lambda \leq {4 \sqrt{2} \pi \over { 3 G_F m_t }} \sim 2.9 ~{\rm TeV}
{}~~~. } \eqno(1)$$
\no Since this energy scale is around $4 \pi v \sim 3.1$ TeV[5][6], the
 scale of dynamical electroweak symmetry breaking,
 it suggests a possible connection between the
top quark mass and the Fermi scale.

One interesting observation, based on
the fact that the top quark Yukawa
coupling is of order one, is that
 the top quark may be compared to
the constituent quarks in QCD, whose mass ratios
to the
pion decay constant is also a number of order unity.
Since the constituent quarks get masses dynamically,
  one would
expect logically that the top quark also gets its mass dynamically.

The above argument not only makes a heavy top
quark natural, but also
results in a lot of interesting phenomena associated with the top quark.
As an example, one
expects that the axial coupling constant of the top quark to
the neutral gauge boson, $g_A$, will deviate from the standard model
because the similar quantity for the constituent quark,
$G_A$, is renormalized[6].
These predictions can be tested in the future Colliders.
And if found experimentally to be true, it will provide a real clue
 to the mechanism
of the breakdown in the electroweak theory. Note that in QCD
chiral symmetry is broken dynamically by quark condensation.
We should mention that if the top quark has different properties
from the standard model prediction, it will affect experimental
quantities associated with the bottom quark. Indeed,
 an anomalous $\Delta g_A$ at the level of the order $\Delta G_A
\simeq 0.25$ for
the constituent quarks can make the discrepancy of the $Z
\rightarrow b {\overline{b}}$
width measured at LEP with the standard model expectation
disappear[7].

 Furthermore, the constituent quarks are argued to have a non-trivial
structure[8], {\it i.e.}, consisting of a valence quark plus many
quark-antiquark pairs. D.B. Kaplan has shown that the
constituent quarks
can be considered as color solitons ( Kaplan names them
``qualitons" )[9].
In this picture, the constituent quark Q would look like a
current quark q carrying around with it a significant deformation of
the $< {\overline{q}} q>$ background.
The spin, color and baryon number of the constituent quark are topologically
induced and not localized at a point.

In this brief report, we suggest that the top quark
 is a soliton. For simplicity, we
consider here the basic properties of the soliton, such as mass, radius,
 baryon number and its color and spin representation\footnote{[F.1]}{
The flavor quantum number of the soliton in
the coset space $SU(2) \times U(1) / U(1)$ has been
worked out by C. Arnade and J. Bagger[10].}.
 The model we use for the
top quark soliton\footnote{[F.2]}{
This model is similar to that considered by Kaplan[9] for qualitons.}
 is based on a non-linear
realization of the
$SU_L(3) \times SU_R(3) / SU_{L+R}(3)$. The unbroken group
$SU_{L+R}(3)$ is the color gauge group
$SU_c(3)$, so the Goldstone bosons
$\pi^a$
are in the adjoint representation
of $SU_c(3)$. The effective action describing the soliton solution is
given by,

$${\eqalign{
S= & {F_\pi^2 \over 16} \int d^4x ~~ Tr( \partial_\mu U^\dagger \partial^\mu U
                               )  \cr
  & + {1 \over 32 e^2 } \int d^4x ~~ Tr{( [ \partial_\mu U U^\dagger, ~
                          \partial_\nu U U^\dagger ] )}^2 \cr
  & + n ~~\Gamma_{WZ} \cr
  & + {m_\pi^2 F_\pi^2 \over 8} \int d^4x ~ Tr( U - 3 ) ~~, \cr
} }\eqno(2)$$

\no where $U= {\rm exp}( 2 i \lambda^a {\pi^a \over F_\pi} )$,
and $m_\pi$ is the Goldstone boson mass.
It should be pointed out that
we have neglected in (2)
the operators with more powers of derivatives
suppressed by powers of the symmetry breaking scale $\sim 4 \pi v
\sim O(1 ~{\rm TeV})$. The colored Goldstone bosons $\pi^a$ can decay
into
two gluons and the top quark pairs if $m_\pi > 2 m_t$. In this paper, we will
not discuss the phenomena associated with the $\pi^a$ fields, instead we take
the effective action in (2) as a toy model to illustrate some
features
of the top quark as a soliton.

The action in (2) has the same form as that for
a ordinary SU(3) Skyrmion[11]. However, the parameters
 are quantatively different.
In
the original $SU(3)$ Skyrmion model,
$F_\pi \sim O(100 ~{\rm MeV})$ and
$n = N_c = 3$, but
here
$F_\pi \sim v \sim O( 100 ~{\rm GeV} )$ and
$n = 1$. Thus the
soliton will be much heavier than the ordinary Skyrmion and its
quantum numbers will also be completely
different from that of the $SU(3)$ Skyrmion. In the Skyrmion model,
the baryon number of the soliton is one, and
 the lowest baryonic state is in the octet
representation of $SU_{L+R}(3)$. Instead, we have here
that the baryon number is $1\over 3$,
 and the lowest baryonic state is in
 the fundamental representation of $SU_{L+R}(3)$ ( for a detailed
 discussion,
see ref.[9] ). Thus
 the soliton carries
 the quantum number of the top quark\footnote{[F.3]}{
The soliton considered by
D' Hoker and Farhi[12] in the effective action
 generated by integrating out
the heavy quark of the standard model is a color neutral object with
integer baryon number.}.

The technique of studying the
properties of the soliton is well-known.
 First, take the ``hedgehog" Ansatz for $U$:
$${
U_0 = {\rm exp} \{ i F(r) {\hat {\vec r}} \cdot {\vec \tau} \}
{}~~~~~~~, }\eqno(3)$$

\no where the $\tau^i$ are the Pauli matrices embedden in the
color $SU_{L+R}(3)$. In this Ansatz,

$${
F(0) = \pi ~, ~~~~~~~~~ F( \infty ) = 0 ~~. } \eqno(4)$$

\no The energy of the solution (3) is given by

$${
M = M_{cl}[F] + m_{cl}[F] ~~~~~,
    }\eqno(5)$$

\no where

$${
\eqalign{
M_{cl}[F] = & 4 \pi {F_\pi \over e} \int_0^\infty dx [ ~~ {x^2 \over 8}
                  {( {dF \over dx} )}^2 + {\sin^2 F \over 4} \cr
            & + {\sin^4 F \over 2 x^2 } + {( { dF \over dx })}^2 ~~ ]
      ,     \cr }
   } \eqno(6.a)$$

$${
m_{cl}[F] = {m_\pi^2 \pi \over  e^3 F_\pi }
                  ~ \int_0^\infty x^2~ dx~ [ 1 - \cos F ]
 ~~~~, } \eqno(6.b)$$

\no where $x = e F_\pi r$.

The soliton state can be constructed by using the standard quantization
 method[13]. For our purpose, we just give the expression of the
Hamiltonian for the soliton,

$${
H= M + {1\over 2} ( {1 \over I_A}  - {1\over I_B} ) j (j+1)
      + {1 \over 2 I_B} ( C_2 - {1 \over 12} )
  ~~~~~,  }\eqno(7)$$

\no where $j$ is the spin and
$C_2$ is the color
 $SU_c(3)$ Casimir.
In (7),
$${
I_A = {2 \pi \over {3 e^3 F_\pi} } \int_0^\infty
           dx~ \sin^2 F ~[ ~ x^2 + 4 ( \sin^2 F + x^2 {({ dF \over dx })}^2
                    ) ] ~~~,} \eqno(8.a)$$

$${
I_B = {\pi \over {e^3 F_\pi } } \int_0^\infty dx~
                  ~ \sin^2{F\over 2}
                 [ x^2 + 2 \sin^2 F + x^2 {( {dF \over dx })}^2 ]
{}~~~~.  }   \eqno(8.b)$$

\no For a spin $1\over 2$, color triplet soliton,
$j= {1 \over 2}$ and
$C_2 = {4 \over 3}$. Then the mass of the soliton is given by

$${
M_{(soliton)} = M + {3 \over {8 I_A} } + {1 \over {4 I_B} }
{}~~~~~. }\eqno(9)$$

 The isoscalar
 radius of the soliton is given by
$${
r_0 = {< r^2 > }^{1/2}_{I=0}
            = {1 \over {e F_\pi} }
 { \{  { {-2 \over \pi} \int_0^\infty dx ~ x^2 ~ \sin^2 F ~{( { dF \over dx })}
               } \} }^{1/2}
 ~~~~. }   \eqno(10)$$

Now let us calculate $M_{(soliton)}$ and
$r_0$ numerically.
 We take a variational approach in ref.[14] and
assume that $F(r)$ has the following form:

$${
F(x) = ~ 2 ~arctan[ {({x_0 \over x})}^2 ]
{}~~~~~, }  \eqno(11)$$

\no where $x_0$, the soliton size, is the variational parameter. For
simplicity, we consider first the case where $m_\pi = 0$, then

$${
M = {F_\pi \over e} \pi^2 {3 {\sqrt 2} \over 16 }
             ( 4 x_0 + {15 \over x_0 } )
 ~~~~~; } \eqno(12.a)$$

$${
I_A = {1 \over {e^3 F_\pi } } \pi^2 { {\sqrt 2} \over 12 }
            (  6  x_0^3 + 25 x_0 )
{}~~~~~~~;  }  \eqno(12.b)$$

$${
I_B = {1\over { e^3 F_\pi } } \pi^2 { {\sqrt 2} \over 16 }
                   (  4 x_0^3 + 9 x_0  )
{}~~~~~~.  }  \eqno(12.c) $$

\no The $x_0$ can be estimated by minimizing (12.a) with respect to
$x_0$, which gives $x_0 = {\sqrt {15/4}}$. Then we have,

$${
M_{(soliton)} \simeq F_\pi ~ ( {40.544 \over e} +  e^3 \times 0.007 )
{}~~~~~,  } \eqno(13)$$

\no and

 $${
r_0 \simeq {2.19 \over {e F_\pi } }
{}~~~~~~. } \eqno(14)$$
 \no Following Ref.[15], let us consider a quantity $M_{(soliton)} ~r_0$,

$${
M_{(soliton)} ~r_0 \simeq {2.19} \times ( {40.544 \over e^2} +
     0.007 \times e^2 )
{}~~~. } \eqno(15)$$
\no Since $M_{(soliton)}~ r_0$ depends only on one parameter $e$, we would
 be able to
find out a possible minimum value of
the $M_{(soliton)} ~r_0$ by
minimizing (15) with respect to $e$.
 It gives that
$${
M_{(soliton)} ~r_0 \geq 2.333
{}~~~~~, } \eqno(16)$$

\no which is comparable to 2.52[15], Keaton's numerical value.
  One can see that for a top quark $ m_t \sim
   ~ 174$ GeV,
we have
 $r_0 ~ \sim {1\over v}$. It should be pointed out that the minimum value
of eq.(16) corresponds to $e^2=76.11$. In such a situation the quantum and
the classical mass term in eq.(13) are exactly equal and the collective
quantization, as carried out for
the soliton here, is not well justified.
So a smaller $e^2$ should be choosen. Consequently,
  $M_{(soliton)}r_0$ is expected to be
larger than $2.333$. For example, If one takes $e^2=50$, the quantum
 mass is only
about half of the classical mass and now $M_{(soliton)} r_0$
is about $2.542$, which is slightly larger than the minimum value $2.333$.

In summary, we have shown that top quark can be described by a soliton.
The spin, color and the baryon number
of the top quark are topologically
induced and its radius is of order $\sim {1 \over v}$.
The top quark field is not localized at a point and
  will have color form factors. These new physics effects may show up
experimentally as excessive or anomalous top
 production rates and distributions at the Hadron Colliders[16].

Before concluding, we would like to point out that
in order to construct the color topological solitons, we have made use of the
 chiral color
symmetry $SU_L(3) \times SU_R(3)$. Whether
it exists or not
depends on the dynamics of the fundamental theory of the top quark.
As an existence proof, let us consider
the gauge version of
the top quark condensation theory[17], where new strong physics of the top
quark at the $\sim 1$ TeV scale is introduced[18].
 In a specific model proposed by
  Lindner and
Ross[19], the four fermion interaction responsible for
the top quark condensation is,
$${
{\cal L}_{eff} \sim {\overline T_L}\gamma_\mu T_L ~~ {\overline t_R} \gamma^\mu
                     t_R ~~~~~, }\eqno(17)$$
\no where $T_L$ stands for the left-handed top and bottom quark
doublet,
$t_R$ the right-handed top quark.
In (17) the chiral color symmetry does exist[20].

\filbreak

\bb
I thank W. Bardeen, Gil Gat,
S. Nussinov for discussions and B.-L. Young and K. Whisnant
for reading the manuscript.
This work is supported in part by the Office of High Energy
and Nuclear Physics of the U.S. Department of Energy (Grant No.
DE-FG02-94ER40817).

\bb
\ce {\bf References}

\item{[1]}CDF Collaboration, F. Abe et al, FERMILAB-PUB-94/097-E,
(1994).

\item{[2]}R.D. Peccei and X. Zhang, Nucl Phys. {\bf B337}, 269 (1990).

\item{[3]}T. Appelquist and M.S. Chanowitz, Phys. Rev. Lett. {\bf 59},
         2405 (1987).

\item{[4]}W. Marciano, G. Valencia and S. Willenbrock, Phys. ReV. {\bf D40},
         1725 (1989).

\item{[5]}S. Weinberg, Physica {\bf 96A}, 327 (1979).

\item{[6]}H. Georgi and
A. Manohar, Nucl. Phys. {\bf B 234}, 189 (1984).

\item{[7]}X. Zhang, Mod. Phys. Lett. {\bf A9}, 1955 (1994); and
references therein.

\item{[8]}For example, see, H. Fritzsch, CERN-TH. 7079/93,
November 1993.

\item{[9]}D.B. Kaplan, Phys. Lett. B235, 163 (1990); Nucl. Phys. B351,
              137 (1991).

\item{[10]}J. Bagger, (Private communication).

\item{[11]}For a review, see I. Zahed
 and G.E. Brown, Phys. Rept. V142, 1 (1986), and
 references therein.

\item{[12]}E. D'Hoker and E. Farhi, Phys. Lett. B134, 86 (1984);
           Nucl. Phys. B241, 109 (1984).

\item{[13]}G.S. Adkins, C.R. Nappi and E. Witten, Nucl. Phys. B228, 552
 (1983); G.S. Adkins and C.R. Nappi, Nucl. Phys. B233, 109 (1984).

\item{[14]}M. Praszalowicz, Ph. D thesis, Jagellonian University, Poland,
   TPJU -12/91, may 29, 1991.

\item{[15]}G. Keaton, Nucl. Phys. {\bf B425}, 595 (1994); see also,
G. Gomelski, M. Karliner and S. Selipsky, Phys. Lett. B323,
          182 (1994).

\item{[16]}For a phenomenological discussion of the new physics effect
on top quark production, see, for example,
D. Atwood, A. Aeppli and A. Soni, Phys. Rev. Lett. 69, 2754 (1992);
          C. Hill and S. Parke, Phys. Rev. D49, 4454 (1994).

\item{[17]}Y. Nambu, Enrico Fermi Institute Preprint (1989);
V. Miransky, M. Tanabashi, and K. Yamawaki, Phys. Lett.
B221, 177 (1989);
W. Marciano, Phys. Rev. Lett. 62, 2793 (1989);
W. Bardeen, C. Hill and M. Lindner, Phys. Rev. D41, 1647 (1990).

\item{[18]}C.T. Hill, Phys. Lett. {\bf B266}, 419 (1991).

\item{[19]}M. Lindner and D. Ross, Nucl. Phys. B370, 30 (1992).

\item{[20]}
 In the real world, the
chiral color symmetry is explicitly broken, at least
``weakly" (note that
the QCD coupling constant at the electroweak
symmetry breaking scale is small), by QCD gauge interaction.
So non-vanishing Goldstone boson mass in (2) will cause the results
of the variational approach in
 (11) less reliable[14]. However
 the lower limit in (16) is still valid.

\bye